\documentclass[intlimits,twoside,a4paper]{article}
\usepackage[cp1251]{inputenc}

\usepackage[eqsecnum]{cmpj3}

\usepackage{bm}

\font\bba=msbm10 
\font\bbb=msbm8 
\font\bbc=msbm6 
\newfam\bbfam
\textfont\bbfam=\bba
\scriptfont\bbfam=\bbb
\scriptscriptfont\bbfam=\bbc
\def\bb{\fam\bbfam\bba}
\def\R{{\bb R}}

\def\G{{\bb G}}

\issue{2020}{23}{2}{23603}
\doinumber{10.5488/CMP.23.23603}
\title[A symplectic integrator for molecular dynamics on a hypersphere]{A symplectic integrator for molecular dynamics on a hypersphere}
\author[J.-M. Caillol]{J.-M. Caillol}
\address{
Universit\'e Paris-Saclay, CNRS/IN2P3, IJCLab, 91405 Orsay, France}

\date{Received February 11, 2020, in final form March 3, 2020}
\begin{document}

\maketitle

\begin{abstract}
We present a reversible and symplectic algorithm called ROLL, for integrating the equations of motion in molecular dynamics simulations of simple fluids  on a hypersphere $\mathcal{S}^d$ of arbitrary dimension $d$. It is derived in the framework of geometric algebra and shown to be mathematically equivalent to algorithm RATTLE. 
An application to  molecular dynamics simulation of the one component plasma is briefly discussed.

\end{abstract}
\section{Introduction}
It is a great pleasure for me to contribute to this special issue of Condensed Matter Physics dedicated to my colleague and friend \mbox{I. Mryglod}, on the occasion of  his sixtieth birthday. I hope that this
paper, which deals with the theory of molecular dynamics  (MD) simulations, will be of some interest to him.

The idea of using the two dimensional (2D) surface of an ordinary  sphere, i.e.,  space 
$\mathcal{S}^{2}$,  to perform Monte Carlo (MC) and/or MD
simulations of a  2D fluid  can be tracked back to a paper 
by \mbox{J.-P. Hansen {et al.}} devoted to a study of the electron gas at 
the surface of liquid Helium \cite{JPH}. The generalization to MC simulations of 3D systems, implying the use of the 3D surface $\mathcal{S}^3$ of a 4D-hypersphere, is due to Caillol and Levesque \cite{Orsay-1} in their study of 3D ionic liquids.

Recently, a symplectic integrator for MD simulations of 2D systems on 2-spheres was proposed in references~\cite{Lee,Vest}. In this article, we extend this integrator to 3D systems. This can be elegantly done  in the framework of geometric algebra (GA)~\cite{Hestenes-0,Hestenes-1,Hestenes-2,Mac-0,Mac-1,Doran}.

The paper is organized as follows. In section~\ref{SecII}
we discuss the extension of the laws of classical mechanics in an euclidean space of arbitrary dimension
$d$, with an emphasis on the correct definition of the angular momentum. In section~\ref{SecIII} we consider the motion of classical point particles confined on the surface of a hypersphere from the point of view of GA formalism. GA is briefly described in appendix~\ref{geometric-algebra}. We derive the algorithm ROLL in section~\ref{SecIV}. ROLL is shown to be equivalent to algorithm RATTLE~\cite{RATTLE} in appendix~\ref{b} from which it inherits all its properties (reversibility and symplecticity). We then present an application of ROLL to MD simulations of the one component plasma (OCP)
in section~\ref{OCP}. We conclude in section~\ref{Conclusion}.
\section{Classical mechanics in an euclidean space of arbitrary dimension}\label{SecII}
\subsection{General discussion}\label{general}
Let us first fix some definitions and notations. We associate to the usual inner product 
space $\R^d$ the euclidean affine space $E^{d}$, defined as a set of points $\mathrm{M}$ such that, given an origin $\mathrm{O}$, there is a unique vector $q \in \R^d$ such that $\overrightarrow{\mathrm{OM}}=q$. Once
$\mathrm{O}$ is fixed, we identify the two spaces $E^d$ and $\R^d$. We denote by $\lbrace e_i\rbrace $, $i=1, \ldots, n$, the standard orthonormal basis of $E^{d}$ with the property that $e_i \cdot e_j = \delta_{ij}$, where ``$\cdot$'' is the usual scalar product of $\R^d$ and $\delta_{ij}$ the Kronecker symbol.

Extending the laws of classical mechanics to a space $E^{d}$ of arbitrary dimension $d>3$ causes no problem, except for those quantities or laws involving the cross-product of two vectors $a$ and $b$, as the angular momentum, for instance. In physical space $E^3$, the cross-product $a\times b$ is an axial vector, perpendicular to the 2-plane $\mathrm{span}(a,b)$ and of a magnitude equal to the area of the parallelogram with edges $a$ and $b$. This hybrid construction cannot be generalized to dimensions $d>3$ since there is not one, but $(d-2)$ directions orthogonal to the 2-plane $\mathrm{span}(a,b)$.
In order to tackle these difficulties, we introduce the geometric algebra  $\G^d$ associated to $\R^d$.\footnote{GA is a relatively new concept, with a lot of applications in many branches of physics, robotics and engineering. 
$\G^d$ is a Clifford algebra, the elements of which are  called multi-vectors which represent  subspaces of $\R^n$. The reader not acquainted with this subject will find a digest in appendix~\ref{geometric-algebra} and much more details in the recent review by A. Macdonald~\cite{Mac-0} and his remarkably clear elementary textbook~\cite{Mac-1}. At a more advanced level, the reader should consult references~\cite{Hestenes-0,Hestenes-2,Doran}. For applications to physics, the books~\cite{Hestenes-1,Doran} will be consulted with profit. We adopt the notations and definitions of Macdonald throughout this
paper.}

In dimension $d=3$, we have  $a\times b= \left( a \wedge b \right)^{\star}$, where  $\star$ denotes the dual~\eqref{def-dua} of the outer product  $a \wedge b$. Actually, the outer product $a \wedge b$ is a bivector representing the plane $\mathrm{span}(a,b)$, \emph{not a vector}, whose magnitude is equal to the area of the parallelogram with edges $a$ and $b$ (cf. the definition~\eqref{fonda} in appendix~\ref{geometric-algebra}). The point is that this definition of the outer product $a \wedge b$ can be extended to an algebra $\G^d$ of arbitrary  dimension $d$, in contrast to the cross-product which is defined only for $d=3$.

Thus, for a system of $N$ point particles (positions $q_{\alpha}$, linear momenta
$p_{\alpha}=m_{\alpha} \dot{q}_{\alpha}$ (as usual, dots denote time derivative), $\alpha=1, \ldots, N)$ of $E^d$, we are  led to  define the  angular momentum as~\cite{Hestenes-1,Doran}
\begin{align}
\label{mom-cin}
 \mathrm{K} &= \sum_{\alpha=1}^N q_{\alpha} \wedge p_{\alpha}. 
\end{align}
$\mathrm{K}$ is a bivector of the GA algebra $\G^{d}$. In dimension $d=3$, the dual of $\mathrm{K}$ is the usual 1-vector angular momentum.
\begin{equation}
\label{mom_cin_dual}
 \mathrm{K}^{\star} = \sum_{\alpha=1}^N q_{\alpha} \times p_{\alpha}\, \quad \text{(only for } d=3).
\end{equation}

It follows from $\dot{q}_\alpha \wedge p_{\alpha}=0$  that a well known theorem of classical mechanics~\cite{Gold,Hestenes-1,Doran} is generalized to arbitrary dimension:
\begin{align}
\label{th-mom-cin}
 \frac{\rd}{\rd t}\mathrm{K} &= \sum_{\alpha=1}^N q_{\alpha} \wedge f_{\alpha}\,, 
\end{align}
where $f_{\alpha}$ is the total force acting on particle $\alpha$. Following Goldstein~\cite{Gold} we decompose    $f_{\alpha}$ as the sum of an external force $f_{\alpha}^\text{e}$ and the sum of the pair-wise additive forces due to the other particles $\sum_{\beta \ne \alpha} f_{\beta \alpha}$.
If   Newton's weak law of action and reaction holds, {i.e.}, if
$f_{\beta \alpha}= -f_{\alpha \beta}$,
then we have
\begin{align}
\label{th-mom-cin-bis}
 \frac{\rd}{\rd t}\mathrm{K} &= \sum_{\alpha=1}^N q_{\alpha} \wedge f_{\alpha}^\text{e}  . 
\end{align}

In the case where there are no external forces, the angular momentum bivector
$\mathrm{K}$ is a conserved dynamical variable of the motion. This means the conservation of its $\binom{d}{2}$ scalar components.
As it is well known, this is a consequence of the rotational invariance of the system, as expressed by Noether's theorem~\cite{Gold,Doran}.
As a warm up, let us derive this theorem in the framework of the GA formalism.  

\subsection{Rotational invariance and the conservation of the angular momentum bivector}
Henceforth in this section, we restrict ourselves to systems consisting of $N$ classical point particles with identical masses $m$  in Euclidean space $E^d$.
The Lagrangian  reads:
\begin{align}\label{Lagrangian}
\mathcal{L}\left( q_{\alpha}, \dot{q}_{\alpha}\right) & =
\frac{m}{2} \sum_{\alpha=1}^N \dot{q}_{\alpha}^2         - \frac{1}{2}\sum_{\alpha \neq \beta}
v\left(\vert q_{\alpha} - q_{\beta} \vert  \right) ,
\end{align}
where the pair potential depends only on the distance $ \vert q_{\alpha} - q_{\beta} \vert$ between the particles.
Extending Euler-Lagrange equations to an arbitrary dimension $d$ causes no trouble, with the well known result~\cite{Gold}
\begin{align}\label{Lagrange-eq}
\frac{\rd}{\rd t}\frac{\partial \mathcal{L}}{\partial \dot{q}_{\alpha}} & =\frac{\partial \mathcal{L}}{\partial q_{\alpha}} \quad \text{   for } \quad \alpha=1, \ldots, N.
\end{align}

Now we apply  Noether theorem to  the Lagrangian function~\eqref{Lagrangian}. Clearly, 
$\mathcal{L}\left( q_{\alpha}, \dot{q}_{\alpha}\right)$
is invariant under any rotation of $\mathrm{O}(d)$  with center 
$\mathrm{O}$. An elemental rotation of space $\textrm{E}^{d}$ 
is described firstly by its invariant plane, containing the origin, and represented by some unit bivector $\ri$ (with $\ri^2 = -1$), and secondly, by a (scalar) magnitude $\theta$. Bivector $\Omega = \ri \theta$ is called the angle of rotation.  The action of this rotation is described by a member of the geometric algebra $\G^{d}$, $R$, called a rotor (or a generalized complex number) defined as  \cite{Mac-0,Mac-1,Doran,Thomas}
\begin{align}
\label{rotor}
R = & \exp \left( -\frac{\Omega}{2} \right) = \cos \frac{\theta}{2}- \sin \frac{\theta}{2} \ri.
\end{align}
Under rotation, every element $A$ of the algebra is transformed according to the law 
\begin{align}
\label{act_rotor}
A&= R A R^{-1} \ ,
\end{align}
where the inverse of $R$, $R^{-1}=R^{\dagger}=\exp\left( +\Omega/2 \right)$.
Therefore, for an infinitesimal rotation $\delta \Omega$, the variation of $q_{\alpha}$ is given by
\begin{align}
\label{var_1}
\delta q_{\alpha} &= \frac{1}{2}\left(  
                         q_{\alpha} \delta \Omega  - \delta \Omega q_{\alpha}  \right)
                         +\mathcal{O} \left( \langle \delta \Omega^2  \rangle_1\right) 
                  =  q_{\alpha} \cdot \delta \Omega 
                  +\mathcal{O} \left( \langle \delta \Omega^2  \rangle_1\right) \, ,      
\end{align}
where the medium-dot ``$\cdot$'' denotes the inner product of the algebra,
cf. equation~\eqref{inner}. {Idem}, the
variations of the velocities are given by
\begin{align}
\label{var_2}
\delta \dot{q}_{\alpha} &= \dot{q}_{\alpha} \cdot \delta \Omega 
                    +\mathcal{O} \left( \langle \delta \Omega^2  \rangle_1\right) \,  .  
\end{align}
The infinitesimal rotation yields the following  variation for
 the Lagrangian 
\begin{align}
\label{hh}
\delta \mathcal{L} & = \sum_{\alpha=1}^N
 \left[ \frac{\partial \mathcal{L}}{\partial q_{\alpha}}
 \cdot \left( q_{\alpha} \cdot  \delta \Omega \right)
 +\frac{\partial \mathcal{L}}{\partial \dot{q}_{\alpha}} 
 \cdot  \left( \dot{q}_{\alpha} \cdot \delta \Omega \right)
  \right] ,
\end{align}
where we took care of the non-associativity of the inner product. However, by taking into account Euler-Lagrange equations
as well as the identity~\eqref{util} of appendix~\ref{geometric-algebra},
equation~\eqref{hh} can be recast as 
\begin{align}
\label{var_L}
\delta \mathcal{L} &= \left( \frac{\rd}{\rd t}\sum_{\alpha=1}^N p_{\alpha} \wedge q_{\alpha} \right)
\cdot \delta \Omega \, ,
\end{align}
where we introduced  momentum $p_{\alpha}
= \partial \mathcal{L}/\partial \dot{q}_{\alpha}=m \dot{q}_{\alpha}$.
Requiring that $\delta \mathcal{L}= 0$ for an arbitrary infinitesimal bivector 
$\delta \Omega$  yields, as expected, the conservation of the total angular momentum $\mathrm{K}$:
\begin{align}
\label{conserv_mom_cin}
 \frac{\rd}{\rd t}\sum_{\alpha=1}^N q_{\alpha} \wedge p_{\alpha} &=0 .
\end{align}
\section{Classical mechanics on a hypersphere}\label{SecIII}
\subsection{The bivector angular velocity}\label{config}
The hypersphere $\mathcal{S}^d(\mathrm{O,R})$ of center $\mathrm{O}$ and radius $\mathrm{R}$, {i.e.}, the set of points $\mathrm{M}$ of $E^{d+1}$ such that 
$\overrightarrow{\mathrm{OM}}^2=q^2=R^2$,  may be seen  as a $d$-dimensional vector manifold  of the euclidean space $E^{d+1}$~\cite{Hestenes-2}. 
It is convenient to introduce  the unit hypersphere
$\mathcal{S}^d=\mathcal{S}^d(\mathrm{O,R=1})$ of unit vectors $\xi=q/R$.
The space $E^{d}_{\xi}$ tangent to $\mathcal{S}^d(\mathrm{O,R})$ at point $q=R \xi$ plays a central role, here  the particle velocities live. It is an euclidean hyperplane of dimension $d$, spanned by the $d$ orthonormal vectors $\lbrace e_i(\xi)\rbrace$, $i=1, \ldots, d$.  When complemented with vector $\xi$, this system of vectors forms a local orthonormal basis of full space $E^{d+1}$. 
$E^{d}_{\xi}$ is  represented by its unit pseudo-scalar
$\mathbf{I}^{d}({\xi}) =e_1(\xi) e_2(\xi) \ldots e_n(\xi)$~\cite{Hestenes-2,Doran}. Vector $\xi$ and the  d-blade
$\mathbf{I}^{d}({\xi})$ are related by a duality relation~\eqref{toto}, hence
$\xi^{\star}=\pm \mathbf{I}^{d}({\xi})$, the sign depending on the orientation of the local basis of $E^{d+1}$. Therefore, the projection of
any blade $A$ of $\G^{d+1}$ onto $E^{d}_{\xi}$ is given
by~\eqref{proj-lam}
\begin{align}
\label{Proj-0}
\mathrm{P}_{\xi^{\star}}\left( A \right)&=(A \cdot \xi^{\star}) \left( \xi^{\star} \right)^{-1}, 
\end{align}
which gives, in the case of a vector $a\in E^{d+1}$ 
\begin{align}
\label{Proj}
\mathrm{P}_{\xi^{\star}}\left( a \right)&=
 a - \left(\xi \cdot a  \right) \xi  .
\end{align}

Let us consider the dynamics of a single point particle $q$ on 
$\mathcal{S}^d(\mathrm{O,R})$. Its dynamics is that of a rotator of 
$E^{d+1}$: its position at time $t$ can be obtained from that at time $t=0$
by a rotation of $\mathrm{O}(d+1)$ with the center~$\mathrm{O}$ described by some rotor $R(t)$. 
 Thus, we have~\cite{Hestenes-1,Doran}
\begin{align} \label{rotation}
\xi(t)= R(t) \xi_0 R^\dagger (t)  ,
\end{align}
where $R^\dagger (t) = R^{-1} (t)$ is the reverse of 
$R(t)$~\cite{Mac-1,Hestenes-1,Doran}. Taking the time derivative of 
equation~\eqref{rotation}, one finds that
\begin{align} \label{rotation_bis}
\dot{\xi} & = \xi \cdot \Omega \, , 
\end{align}
 where the  angular velocity  $\Omega(t)$ is a bivector defined by
\begin{align} \label{omega}
\Omega    & =-2 \dot{R} R^\dagger  .
\end{align}
At this point it must be stressed that equation~\eqref{rotation} 
does not fully  determine  the rotor $R(t)$ [and thus $\Omega(t)$]. 
Recall that an elemental rotation of $\mathrm{O}(d+1)$ is defined by 
a 2-plane and an amplitude. 
If this plane is a subspace of the tangent hyperplane $\xi^{\star}$,
then this rotation leaves vector $\xi$ unchanged. 
Therefore, there is some freedom and, in order to fix it,
we decompose $\Omega$ upon the standard basis of the bivectors of the algebra
\begin{subequations}
\begin{align}
\Omega   & = \Omega^{\perp} + \Omega^{\parallel}   \, , \\
\Omega^{\perp} & = \mathrm{P}_{\xi^{\star}}\left( \Omega \right) = \sum_{1 \leqslant \alpha < \beta \leqslant d} \Omega^{\perp}_{\alpha \beta} \ e_{\alpha}(\xi) \wedge e_{\beta}(\xi) \,, \label{decompo-a} \\
  \Omega^{\parallel} & = \sum_{1 \leqslant \alpha \leqslant d} \Omega^{\parallel}_{\alpha} \ e_{\alpha}(\xi) \wedge \xi  . 
  \label{decompo-b}
 \end{align}
\end{subequations}
One readily obtains from expansion~\eqref{decompo-a} that $\xi \cdot \Omega^{\perp}= \langle \xi \Omega^{\perp} \rangle_1 =0 $ and thus $\xi \cdot \Omega=
\xi \cdot \Omega^{\parallel}$. Similarly, from~\eqref{decompo-b} one deduces that  $\xi \wedge \Omega^{\parallel} =0 $ 
and thus $\xi \wedge \Omega=
\xi \wedge \Omega^{\perp}$. 
It is thus possible to chose $\Omega^{\perp}=0$ which fully 
determines the angular velocity. We can thus impose the condition
\begin{align} \label{condition}
\xi \wedge \Omega =0.
\end{align}
Indeed, from  $\Omega^{\perp}=0$ and 
$\dot{\xi}  = \xi \cdot \Omega $ one obtains all the coefficients of the 
expansions~\eqref{decompo-a} and~\eqref{decompo-b}: $\Omega^{\perp}_{\alpha \beta}=0$ and 
$\Omega^{\parallel}_{\alpha}= -\dot{\xi} \cdot e_{\alpha}(\xi)$, which means that
$\Omega = \xi \wedge \dot{\xi} = \xi  \dot{\xi} $. This expression for $\Omega$  shows that  the two additional conditions  $\dot{\xi} \wedge \Omega=0$ and  $\xi \wedge \dot{\Omega}=0$ also hold.

It is interesting to consider more closely the case $d=2$ for a comparison with the work of Lee {et~al.},~\cite{Lee}. In this case, 
the dual of bivector $\Omega$ is an axial vector $\omega$. And one recovers the usual expression for the vector angular velocity 
$\omega = \Omega^{\star}= \xi  \times \dot{\xi}$ and  the usual relation
$\dot{\xi}=\omega \times \xi$~\cite{Lee}. The condition $\xi \wedge \Omega =0$ becomes
$\xi \cdot \omega =0$.
\subsection{Equation of motion on a hypersphere}
To simplify the notations we consider the case  of a single point particle. For the matters discussed here, the extension to the case of $N$ particles is trivial.
This particle,  of mass $m$, position $q$, is subjected to some external,  unspecified force $f$ and constrained  to move on the surface of the sphere $\mathcal{S}^d\mathrm{(O,R)}$. Again, $\xi = q/R$ denotes its position on the unit hypersphere $\mathcal{S}^d$.

Two points of view can be adopted: the motion can be either described as a motion in $E^{d+1}$ with holonomic constraints or by the theorem~\eqref{th-mom-cin} derived in section~\ref{general} and rederived in section~\ref{varia-continu} using a Lagrangian formalism in the framework of GA.
\subsubsection{Constraints}\label{gonzo}
The particle  is subjected to  the holonomic constraint  
\begin{align}
\label{Contraintes}  
  \sigma(q)   &=  q^2 -R^2 = 0  .
\end{align}
If $f=-\nabla_q V(q)$ the Newton's equation of motion can be obtained as the Euler-Lagrange equation associated with the constrained Lagrangian $\mathcal{L}^{\star}= m \dot{q}^2/2 - V(q) - \lambda  \sigma(q)$~\cite{Gold,SHAKE}
\begin{align}
\label{Newton}
  m \ddot{q}  &= f - 2 \lambda q  .   
\end{align}
The force  due to the constraint  $-\lambda \nabla_q \sigma = -2 \lambda q$ is, of course, perpendicular to the tangent plane $\xi^{\star}$.
Equation~(\ref{Newton}) can be simplified as 
\begin{subequations}
\begin{align}
\label{Newton_barre}
 m \ddot{\xi}  &= \frac{f^{\perp}}{R} - 2 \overline{\lambda} \xi  ,  \\
 \overline{\lambda} &= \lambda - \frac{f \cdot \xi}{2R} \, ,
\end{align}
\end{subequations}
 where $f^{\perp}$ is the projection of the force $f$ in the tangent hyperplane
  $E^d_{\xi}$ [cf. equation~\eqref{Proj}]: 
 \begin{align}
 f^{\perp} & = \mathrm{P}_{\xi^{\star}}(f) =f - (f\cdot \xi) \xi .
 \end{align}
We note that, taking the scalar product of both sides of~\eqref{Newton_barre} with $\xi$ yields
$\overline{\lambda}=(m/2) \dot{\xi}^2$ where we used the ``hidden constraint'' $\xi \cdot \ddot{\xi} + \dot{\xi}^2=0$ obtained by taking twice the time derivative of the constraint equation~\eqref{Contraintes} valid for all times.  $\overline{\lambda}$ has been eliminated and the equation of motion reads as
\begin{align}\label{Newton_I}
m \ddot{\xi}  &= \frac{f^{\perp}}{R} - m \dot{\xi}^2 \xi  .
\end{align}
\subsubsection{Euler-Lagrange equations}\label{varia-continu}
We consider a particle confined on $\mathcal{S}^d(0,R)$ with Lagrangian $\mathcal{L}(q,\dot{q})=  (m/2) \dot{q}^2 - V(q)$ where $V(q)$ is an arbitrary potential. Euler-Lagrange equations are obtained by minimizing the action
\begin{align}
S & = \int_0^T \mathcal{L}(q,\dot{q})\ \rd t  ,
\end{align}
with respect to infinitesimal variations $\delta q(t)$ such that $\delta q(0)= \delta q(T)=0$. As noted in reference~\cite{Lee}, the expression for $\delta q(t)$ must agree with the geometric structure of the configuration manifold, which is that of a Lie group and not that of a linear vector space. Referring to our discussion of section~\ref{config} we are led to consider $\delta q = q \cdot \delta \Theta $ where $\delta \Theta(t)$ is, as the angular velocity $\Omega(t)$, an arbitrary infinitesimal  time-dependent bivector subjected to the condition $q \wedge \delta \Theta = 0$, {i.e.},
$\delta \Theta_{\perp} = 0$ with the 
notations of section~\ref{config}. Then the variation of velocities should read $\delta \dot{q} =\dot{q} \cdot  \delta \Theta(t) + q \cdot \delta \dot{\Theta}$. In the first-order approximation $(q+ \delta q)^2 \sim R^2$ and $(q + \delta q) \cdot (\dot{q} +\delta \dot{q})\sim 0$. Therefore, the first variation of the Lagrangian is  
\begin{align}\label{var-L}
\delta \mathcal{L} & = \frac{\partial \mathcal{L} }{\partial q} \ \cdot \left( q \cdot \delta \Theta \right) +
\frac{\partial \mathcal{L} }{\partial \dot{q}} \ \cdot \left( \dot{q} \cdot \delta \Theta \right) +
\frac{\partial \mathcal{L} }{\partial \dot{q}} \ \cdot \left( q \cdot \delta \dot{\Theta} \right)  
 = \left( q \wedge \frac{ \partial V}{\partial q} \right) \cdot \delta \Theta - m (q \wedge \dot{q} ) \cdot \delta \dot{\Theta}  , 
\end{align}
where we made use of equation~\eqref{util} and of the antisymmetry of the outer product. It follows that the variation of the action is
\begin{align}\label{var-S}
\delta S & =  \int_0^T \rd t \left[  
q \wedge  \frac{ \partial V}{\partial q} +
m \frac{\rd}{\rd t} \left(\dot{q} \wedge q \right)
\right]   \cdot \delta \Theta  ,
\end{align}
where the usual integration by parts on time $t$ and the use of the boundary conditions at $t=0$ and $t=T$ were made.
Equation~\eqref{var-S} can be rewritten in a more transparent way 
\begin{equation}\label{var-SS}
\delta S  = \int_0^T \rd t\left(  
- q \wedge f^{\perp} + \dot{K}
\right)   \cdot \delta \Theta  ,
\end{equation}
where $K=m q \wedge \dot{q}= m R^2 \Omega$ is the angular momentum of the particle.
Requiring that $\delta S=0$ for an arbitrary variation $\delta \Theta(t)$ does not automatically ensures that the bivector in brackets $ A =- q \wedge f^{\perp} + \dot{K} $
in equation~\eqref{var-SS} vanishes.
To see that, we decompose  $A = A_{\perp} + A_{\parallel}$ as we did
in section~\ref{config}, \textit{idem} for $\delta \Theta$.
We then note that  $A \cdot \delta \Theta = \langle A \ \delta \Theta\rangle_0 = A_{\perp} \cdot \delta \Theta_{\perp}  + A_{\parallel} \cdot \delta \Theta_{\parallel}$. 
Since $\delta \Theta_{\perp}=0$, we have simply
$A \cdot \delta \Theta = \langle A_{\parallel} \delta \Theta_{\parallel} \rangle_0 $. Now we expand $A_{\parallel} =\sum_i A_i e_i(\xi) \xi$ and 
$\delta \Theta(t)  = \sum_i \delta \theta_i(t) e_i(\xi) \xi$ where the $\delta \theta_i(t)$ are  $d$ arbitrary functions of time. Then, the stationary condition  says nothing about  $A_{\perp}$  but implies that all the $A_i$ are zero.
Therefore, we have 
\begin{align} \label{ici}
-q \wedge f^{\perp}+ \dot{K} & = \sum_{i \neq j }
 c_{i j } e_i(\xi) e_j(\xi)  ,
\end{align}
where the coefficients $c_{i j}$ are unknown \textit{a priori}. However, since both $ \dot{K} =q \wedge \ddot{q}$ and $q \wedge f^{\perp}$ have a null projection onto the plane $E^d(\xi)$,  all the $c_{i j }$ of equation~\eqref{ici} vanish. 

Thus, the Euler-Lagrange motion equation  is 
\begin{align}
\dot{K} & = m q \wedge \ddot{q} = q \wedge f^{\perp} \, , \nonumber
\end{align}
or, in term of the angular velocity
\begin{align}\label{yup}
\dot{\Omega} & = \frac{1}{m} \xi \wedge f  .
\end{align}
We recover the angular momentum theorem of equation~\eqref{th-mom-cin-bis}. Note that taking the inner product of both sides of the above equation with vector $q$ yields  the Newton equation~\eqref{Newton_I} derived in  section~\ref{gonzo}.

\section{ROLL: an integrator of the equations of motion on a hypersphere}
\label{SecIV}
Let us begin with a short historical digression. The Verlet algorithm  was originally devised to numerically integrate  the  Newton equations of motion
of a system of $N$ interacting point particles in the first MD simulations of simple classical fluids
or plasmas~\cite{Verlet}. The geometry considered in these early works was the $d=3$ Euclidean space $E^3$, within the usual and convenient periodic boundary conditions~\cite{Verlet}. The so-called velocity-Verlet algorithm derived later by Swipe {et al.} is mathematically equivalent to that of Verlet but, numerically is more stable and precise~\cite{V-Vitesse}. 
SHAKE generalizes Verlet integration to systems of point particles with mechanical constraints, for instance molecules and polymers~\cite{SHAKE}, while RATTLE generalizes the velocity-Verlet algorithm for the same class of systems~\cite{SHAKE}. The analysis of Andersen shows that RATTLE and SHAKE are not numerically equivalent~\cite{RATTLE}. It appears that SHAKE fails to consider hidden constraints on velocities which are correctly taken into account by RATTLE.

SHAKE or RATTLE  can both be used to study the dynamics of particles confined on a hypersphere, but we consider only the superior RATTLE in appendix~\ref{b}. 
In next section~\ref{SEC1} we derive ROLL   from Taylor expansions of the equation of 
motion~\eqref{Newton_I} (cf. section~\ref{SEC1}), and, in section~\ref{SEC2} we derive it  from the exact minimization of a discretized action.
\subsection{Taylor expansions}\label{SEC1}
\subsubsection{ROLL-0}\label{ROLL-0}
We first discretize equation of motion~\eqref{Newton_I} and project it on the hyperplane $\xi^{\star}_n$
\begin{align}
\label{xidd}
\ \mathrm{P}_{\xi^{\star}_n}(\ddot{\xi}_n) &= f^{\perp}_n/(m R)  ,
\end{align}
which gives the projection of the acceleration at time $t=n h$, where $h$ denotes the time increment.
This allows one to approximate the positions at discrete times $t_{n \pm 1}$ with Taylor expansions:
\begin{subequations}
\begin{align}
 \mathrm{P}_{\xi^{\star}_n}(\xi_{n+1}) &= + h \dot{\xi}_{n} +
\frac{h^2}{2 m R}\ f^{\perp}_n +\mathcal{O}(h^3)       \label{blabla}     , \\  \mathrm{P}_{\xi^{\star}_n}(\xi_{n-1}) &=- h \dot{\xi}_{n} +
\frac{h^2}{2 m R}\ f^{\perp}_n +\mathcal{O}(h^3)     \label{bleble}       .
\end{align}
\end{subequations}
Following Verlet and adding both equations gives us~\cite{Verlet} 
\begin{align}
\label{ROLL_0_a}
\mathrm{P}_{\xi^{\star}_n}(\xi_{n+1})& = - \mathrm{P}_{\xi^{\star}_n}(\xi_{n-1}) + \frac{h^2}{ m R}\ f^{\perp}_n +
\mathcal{O}(h^4)  .
\end{align}
It is important to note that the knowledge of the projection  
$\mathrm{P}_{\xi^{\star}_n}(\xi_{n+1}) $ implies the full knowledge of the new position
$\xi_{n+1}$ of the particle at time $t_{n+1}$.
 From equation~\eqref{Proj} we have
\begin{align}\label{eq2}
\xi_{n+1} &= \mathrm{P}_{\xi^{\star}_n}\left( \xi_{n+1}\right) + \alpha \xi_n  .
\end{align}
In order to satisfy the \emph{position constraint} $\xi_{n+1}^2 =  1$ at time $t_{n+1}$,
the unknown parameter $\alpha$ must be:
\begin{align}\label{eq3}
\alpha &=  \pm \sqrt{1 -\mathrm{P}_{\xi^{\star}_n}\left( \xi_{n+1}\right)^2 }  .
\end{align}
For a  small time increment $h$, $\xi_{n} \cdot \xi_{n+1} > 0$ and the positive sign in the r.h.s. of~\eqref{eq3} should be retained.
Therefore,
\begin{align}\label{tres-utile}
\xi_{n+1} &= \mathrm{P}_{\xi^{\star}_n}\left( \xi_{n+1}\right) + \sqrt{1 -\mathrm{P}_{\xi^{\star}_n}\left( \xi_{n+1}\right)^2 } \xi_n  .
\end{align}

Just as the Verlet algorithm, the integrator~\eqref{ROLL_0_a} says nothing about the velocities at time $t_{n+1}$, but generates a sequence of positions $\xi_n$ on the hypersphere. As proposed by Verlet, velocities at time $t_n$ can be obtained by subtracting equations~\eqref{bleble} and~\eqref{blabla} yielding
\begin{align}
\label{ROLL_0_c}
\dot{\xi}_{n}  & = \frac{\mathrm{P}_{\xi^{\star}_n}(\xi_{n+1}) -\mathrm{P}_{\xi^{\star}_n}(\xi_{n-1})}{2h}   .
\end{align} 
Note that these expressions automatically satisfy the hidden constraint on velocities $\xi_{n} \cdot \dot{\xi}_{n} =0$.

To summarize, ROLL-0 is a discrete mapping 
$(\xi_{n-1},\xi_{n}) \rightarrow \xi_{n+1} $ which  can  be written, for a sufficiently small time increment $h$:
\begin{subequations}\label{ALGO-0}
\begin{align}
\mathrm{P}_{\xi^{\star}_n}(\xi_{n+1}) & = -\mathrm{P}_{\xi^{\star}_n}(\xi_{n-1}) + \frac{h^2}{ m R}\ f^{\perp}_n +
\mathcal{O}(h^3)  , \\
\xi_{n+1} &= \mathrm{P}_{\xi^{\star}_n}\left( \xi_{n+1}\right) + 
\sqrt{1 -\mathrm{P}_{\xi^{\star}_n}\left( \xi_{n+1}\right)^2 } \xi_n  \,,  \\
\dot{\xi}_{n}  & = \frac{\mathrm{P}_{\xi^{\star}_n}(\xi_{n+1}) -\mathrm{P}_{\xi^{\star}_n}(\xi_{n-1})}{2h} + \mathcal{O}(h^2) \ ,
\end{align}
\end{subequations}
it satisfies both constraints $\xi_{n+1}^2=1 $ and $\xi_{n} \cdot \dot{\xi}_{n}=0 $.
\subsubsection{ROLL-2} \label{ROLL-2}
We skip ROLL-1 which will be derived in  section~\ref{SEC2} and we present now ROLL-2. We start from a Taylor expansion about time $t_n$ rewritten with the help of bivector angular velocity $\Omega_n$
\begin{align}
\xi_{n+1} & = \xi_{n} + h \ \xi_{n}\cdot \Omega_n + \frac{h^2}{2 m R} f_n +
\mathcal{O}(h^3)\ .
\end{align}
We project the equation  on the tangent plane $E^d_{\xi}$ obtaining
\begin{align}
\mathrm{P}_{\xi^{\star}_n}(\xi_{n+1})& = h \ \xi_{n}\cdot \Omega_n +\frac{h^2}{2 m R}\ f^{\perp}_n +
\mathcal{O}(h^3).
\end{align}
As explained in section~\ref{ROLL-0}, $\mathrm{P}_{\xi^{\star}_n}(\xi_{n+1})$ determines $\xi_{n+1}$.
We need an expression for $\Omega_{n+1}$. For this purpose we discretize equation~\eqref{yup} and adopt the velocity-Verlet recipe:
\begin{align}\label{ROLL_2_c}
\Omega_{n+1} &= \Omega_{n}  + \ \frac{h}{2 m R} \left( 
\xi_{n+1} \wedge f^{\perp}_{n+1} + \xi_{n} \wedge f^{\perp}_{n}
\right)+
\mathcal{O}(h^2).
\end{align}
To summarize, ROLL-2 is a discrete mapping $(\xi_{n},\Omega_{n}) \rightarrow (\xi_{n+1},\Omega_{n+1}) $ which  can  be written, for a sufficiently small time increment $h$:
\begin{subequations}\label{equa-de-ROLL2}
\begin{align}
\mathrm{P}_{\xi^{\star}_n}(\xi_{n+1}) & = h \ \xi_{n}\cdot \Omega_n +\frac{h^2}{ 2 m R}\ f^{\perp}_n +
\mathcal{O}(h^3)\ , \\
\xi_{n+1} &= \mathrm{P}_{\xi^{\star}_n}\left( \xi_{n+1}\right) + 
\sqrt{1 -\mathrm{P}_{\xi^{\star}_n}\left( \xi_{n+1}\right)^2 } \xi_n \ ,  \\
\Omega_{n+1} &= \Omega_{n}  + \ \frac{h}{2 m R} \left( 
\xi_{n+1} \wedge f^{\perp}_{n+1} + \xi_{n} \wedge f^{\perp}_{n}
\right)+
\mathcal{O}(h^2) \ . 
\end{align}
\end{subequations}
\subsection{The discrete action:  ROLL as a variational integrator}\label{SEC2} 
\subsubsection{A variational integrator}\label{varia}
We extend the analyses of Lee {et al.} concerning  variational integrators of the discrete action  on two-spheres~\cite{Lee} to the case of hyperspheres of arbitrary dimension. Viewing Verlet integrator as an exact variational principle of a discrete action rather than the approximate Taylor expansion of exact equations of motion  is an idea due to Marsden and collaborators, see e.g., reference~\cite{Marsden}. Here, it yields a unification of ROLL-0 and ROLL-2, the derivation of ROLL-1, and shows the mathematical equivalence of these three versions of ROLL.

As in section~\ref{SEC1} we consider only the case of one particle in the potential $V(\xi)$
since the extension to $N$ particles is trivial.  The discrete Lagrangian of the particle can be defined as~\cite{Marsden}
\begin{align}
\mathcal{L}_{d_n} & = \frac{mR^2}{2 h} 
\left( \xi_{n+1} - \xi_{n}\right)^2 -
                     \frac{h}{2} \left(V_{n} + V_{n+1}\right),
\end{align}
where $V_n \equiv V(\xi_n)$ is the total potential energy of the system in the configuration $\xi_n$
and at discrete time $t_n=n h$. The discrete action between times $t_0=0$ and $T=n_{\mathrm{max}}h$ is defined as
\begin{align}\label{action}
S & = \sum_{n=0}^{n_{\mathrm{max}}-1}  \mathcal{L}_{d_n}  .
\end{align}
We now compute the variation  $\delta S$ induced by some infinitesimal 
variation of the positions $\delta \xi_n$. As in the continuous case discussed in section~\ref{varia-continu}, we must be careful in the definition of $\delta \xi_n$ and take  account of the Lie group structure of the configurational space. Therefore, at each time $t_n$, we consider an infinitesimal variation in the tangent $n$-plane $\xi_{n}^{\star}$: $\delta \xi_n = \xi_n \cdot \delta \Theta_n $, where 
 $\delta \Theta_n$ is some bivector of the form $\delta \Theta_n \equiv \delta \Theta^{\parallel}_n  = \sum_{1 \leqslant i \leqslant d} \epsilon_i^n \ e_{i}(\xi_n) \wedge \xi_n$ where the scalars
 $\epsilon_i^n$ are $d$ arbitrary infinitesimal variations. As discussed in 
section~\ref{varia-continu}, this expansion of $\delta \Theta_n$ ensures that the projection of bivector $\delta \Theta_n$ in the tangent plane $E^n_{\xi_n}$ is zero, in other words that $\xi_n \wedge \delta \Theta_n =0$.

Discrete Hamilton's principle states that $\delta S =0$ for arbitrary  $\epsilon_i^n$ subject  to the boundary conditions $\epsilon_i^0 = \epsilon_i^{n_{\mathrm{nmax}} - 1}=0$. The variation of the Lagrangian is
\begin{align}\label{dL}
\delta  \mathcal{L}_{d_n} & = \mathrm{D}_{\xi_n}\mathcal{L}_{d_n} \cdot
\delta \xi_n + \mathrm{D}_{\xi_{n+1}}\mathcal{L}_{d_{n}} \cdot
\delta \xi_{n+1} \ ,
\end{align}
where we have used Marsden's notations $\mathrm{D}_{\xi_n}= \partial /\partial \xi_n $ and $\mathrm{D}_{\xi_{n+1}}= \partial /\partial \xi_{n+1} $. It follows from~\eqref{dL} and from a discrete integration by parts, ({i.e.}, a relabeling of the second  sum), that the variation of the action is 
\begin{align}
\delta S &= \sum_{n=1}^{n_{\mathrm{max}}-1}
            \left[ \mathrm{D}_{\xi_n} \mathcal{L}_{d_n} +
            \mathrm{D}_{\xi_n} \mathcal{L}_{d_{n-1}}\right] 
            \cdot \delta \xi_n \,  \nonumber \\
&=  \sum_{n=1}^{n_{\mathrm{max}}-1}\left[
\frac{m R^2}{h} \left( 2 \xi_n - \xi_{k-1} - \xi_{k+1} \right) + h \xi_n f_n^{\perp}
 \right] \cdot\left(  \xi_n \cdot  \delta \Theta_n \right)  \nonumber  \\
&=  \sum_{n=1}^{n_{\mathrm{max}}-1}\left[
\frac{m R^2}{h} \xi_n \wedge \left( \xi_{n-1} + \xi_{n+1} \right) -h \xi_n \wedge f_n^{\perp}
 \right] \cdot  \delta \Theta_n  ,
\end{align}
where we made use of equation~\eqref{util}. The condition $\delta S=0$ for an arbitrary $\delta \Theta$ with the condition $\xi_n \wedge \delta \Theta_n =0$ leads, as discussed at length in section~\ref{varia-continu}, to the equations
\begin{align}\label{final}
\frac{m R^2}{h} \ \xi_n \wedge \left( \xi_{n-1} + \xi_{n+1}\right) &=  \xi_n \wedge  f_n^{\perp} \ .
\end{align}
After taking the inner product of both sides of~\eqref{final}, we obtain
\begin{align}
\label{ROROLL_0_a}
\mathrm{P}_{\xi^{\star}_n}(\xi_{n+1}) & = - \mathrm{P}_{\xi^{\star}_n}(\xi_{n-1}) + \frac{h^2}{ m R}\ f^{\perp}_n \ ,
\end{align}
which is, when complemented with equation~\eqref{tres-utile}, nothing else but the integrator ROLL-0 [cf. equations~\eqref{ALGO-0}].
\subsubsection{Linear and angular momenta}
The linear momentum $p_n$ is defined in the tangent plane $E^d(\xi_n)$ as follows~\cite{Marsden,Lee}:
\begin{align}
p_n \cdot \delta q_n & = R p_n \cdot \delta \xi_n = -\mathrm{D}_{\xi_n}\mathcal{L}_{d_n}  \cdot \delta \xi_n .
\end{align}
This is satisfied for any $\delta \xi_n$ perpendicular to $\xi_n$, {i.e.}, of the form
$\delta \xi_n = \xi_n \cdot \delta \Theta_n $ with $\xi_n \wedge \delta \Theta_n =0$.
Therefore, we have 
\begin{align}
p_n \cdot \left(\xi_n \cdot \delta \Theta_n \right)  &=\frac{1}{R}\left[
\frac{m R^2}{h}\left( \xi_{n+1}-\xi_{n} \right) +\frac{h}{2}\frac{\partial}{\partial \xi_n} V_n 
 \right] \cdot \left( \xi_n \cdot \delta \Theta_n \right)  .
\end{align}
This equation must be satisfied for any $\delta \Theta_n \equiv \delta \Theta_n^{\parallel}$ and, following the same technique as in section~\ref{varia}, we find after some algebra
\begin{subequations}\label{momenta-n}
\begin{align}
p_n      & = \frac{m R}{h} \mathrm{P}_{\xi^{\star}_n}\left(\xi_{n+1} \right)  -\frac{h}{2} f_n^{\perp} \,,  \label{momenta-na} \\
\Omega_n & \equiv \xi_n \wedge p_n/(m R) = \frac{1}{h} \xi_{n} \wedge \xi_{n+1} -\frac{h}{2 m R}\xi_n \wedge f_n^{\perp} \, .\label{momenta-nb}
\end{align}
\end{subequations}
 Similarly, one defines the momentum $p_{n+1}$ as
\begin{align}
p_{n+1} \cdot \delta q_{n+1} & = R p_{n+1} \cdot \delta \xi_{n+1} =
 +\mathrm{D}_{\xi_{n+1}}\mathcal{L}_{d_{n+1}}  \cdot \delta \xi_{n+1} \, ,
\end{align}
and one obtains, with computations in the same vein as those used to compute 
$ p_{n}$, the results
\begin{subequations}\label{momenta-nn}
\begin{align}
p_{n+1}     & = - \frac{m R}{h} \mathrm{P}_{\xi^{\star}_{n+1}}\left(\xi_{n} \right)  +\frac{h}{2} f_{n+1}^{\perp} \,,  \label{momenta-nna} \\
\Omega_{n+1} & = \frac{1}{h} \xi_{n} \wedge \xi_{n+1} +\frac{h}{2 m R} \xi_{n+1} \wedge f_{n+1}^{\perp} \, .\label{momenta-nnb}
\end{align}
\end{subequations}
From equations~\eqref{momenta-n} and~\eqref{momenta-nn} one easily obtains  two versions of ROLL integrator.
\begin{itemize}
\item ROLL-1

From equations~\eqref{momenta-na} and ~\eqref{momenta-nna} one obtains a discrete mapping
$(\xi_{n},p_{n}) \rightarrow (\xi_{n+1},p_{n+1}) $ 
which  can  be written, for a rather small time increment $h$:
\begin{subequations}\label{ROLL-1}
\begin{align}
\mathrm{P}_{\xi^{\star}_n}\left(\xi_{n+1} \right) &= \frac{h}{m R}p_n + 
\frac{h^2}{2 m R}f_{n}^{\perp} \,, \\
\xi_{n+1} &= \mathrm{P}_{\xi^{\star}_n}\left( \xi_{n+1}\right) + 
\sqrt{1 -\mathrm{P}_{\xi^{\star}_n}\left( \xi_{n+1}\right)^2 } \xi_n \, ,  \\
p_{n+1}     & = - \frac{m R}{h} \mathrm{P}_{\xi^{\star}_{n+1}}\left(\xi_{n} \right)  +\frac{h}{2} f_{n+1}^{\perp}  .
\end{align}
\end{subequations}
Equations~(\ref{ROLL-1}) constitute ROLL-1 integrator.
After \emph{defining} $\dot{\xi}_n$ by the identity $p_n \equiv m R \dot{\xi}_n$, it turns out that  they are  identical to  the equations~\eqref{RATTLE-S}, constitutive of RATTLE algorithm, which are derived in appendix~\ref{b}. 
\item ROLL-2

From equations~\eqref{momenta-nb} and ~\eqref{momenta-nnb} one obtains a discrete mapping
$(\xi_{n},\Omega_{n}) \rightarrow (\xi_{n+1},\Omega_{n+1}) $ which is nothing else but ROLL-2, derived in another way in section~\ref{ROLL-2}
[cf. equations~\eqref{equa-de-ROLL2}].
\end{itemize}
\subsubsection{Properties of ROLL}
It follows from the analysis of section~\ref{varia} that ROLL-1 and ROLL-2 are two mathematically equivalent integrators. 
ROLL-1 is also mathematically equivalent to RATTLE. Therefore,  all these algorithms share the same properties and are  reversible and symplectic, as shown for RATTLE in reference~\cite{RATTLE,Sympl}. In the case $d=2$,  it is a simple exercice to show  that ROLL-2 coincides with the integrators of references~\cite{Vest,Lee}.

\section{Numerical application}
\label{OCP}
The main application of ROLL  should be for MD simulations of simple fluids. Actually, the results of preceding sections can  be easily  extended  
to a system of $N$ particles. In case of pairwise additive potentials, for instance, the perpendicular force $f_{\alpha}^{\perp}$ acting on particle ``$\alpha$'' is
\begin{align}
f_{\alpha}^{\perp} &= -\frac{1}{R}\sum_{\beta \ne \alpha}
 \frac{\partial  v(\psi_{\alpha \beta})}{\partial \psi_{\alpha \beta}} \textrm{t}_{\alpha \beta}(\xi_{\alpha}) \,, \label{force}
\end{align}
where it had been recognized that the pair potential $v(\psi_{\alpha \beta})$ 
 depends only on the geodesic distance  $\psi_{\alpha \beta} = \cos^{-1}( \xi_{\alpha} \cdot \xi_{\beta}) $ on the unit hypersphere. In equation~\eqref{force} $\textrm{t}_{\alpha \beta}(\xi_{\alpha})$ denotes the tangent to the geodesic $ \widehat{(\xi_{\alpha},\xi_{\beta}})$ at point $\xi_{\alpha}$ and
orientated from $\xi_{\alpha}$ to  $\xi_{\beta}$~\cite{cl,c1}.

We applied  the ROLL algorithm  to MD simulations of the three-dimensional version of the  OCP. The OCP is a model which consists of identical point charges $q$ of mass $m$, immersed in a uniform neutralizing background of charge density $\rho=-nq$ where $n=N/V $ is the number density of particles ($V=2 \piup^2 R^3$ volume of the system)~\cite{Brush,Baus}.
In the thermodynamic limit,
its properties depend only on the  dimensionless 
coupling parameter 
$\Gamma=\beta q^{2}/a$ [$\beta=1/k T$, $T$ temperature, $k$ Boltzmann constant,
and $a$ ion-sphere radius,  defined in $E^3$, by $a=(4\piup n/3)^{-1/3}$]. 

The expression for the  configurational  energy of the OCP  in ${\cal S}^{3}(\mathrm{O,R})$ was derived in~\cite{Caillol1,Caillol2} and will not be repeated here. We only mention that the pair  potential  $v(\psi)$ entering the configurational energy is essentially the electrostatic potential created by a  point charge embedded in its
neutralizing  background. It is obtained by solving
Poisson's equation analytically in 
${\cal S}^{3}(\mathrm{O,R})$~\cite{cl,c1}. 

In reference~\cite{Caillol1,Caillol2}, MC simulations
in the canonical ensemble, performed on the hypersphere $\mathcal{S}^3(\mathrm{O,R})$, produced very accurate results for the thermodynamic limit of the internal energy of the model with a relative precision $p\sim 10^{-5}$ in the range $1 \leqslant \Gamma \leqslant 190$. In these MC simulations, the parameter $\Gamma$ is fixed while, in the MD simulations, the total energy is fixed and the potential and kinetic energy (and therefore, parameter $\Gamma$) must both be computed. 

In our simulations, we have chosen a system of units such that $q=1$, $m=1$,
and $a=1$. With this choice, the unit of time, for instance is
$t_0= (m a^3/q^2)^{1/2}=1$.
We computed the potential energy $V$ and the  kinetic energy $KIN = (m/2) R^2\sum_{\alpha} \dot{\xi}_{\alpha}^2$ as well as their sum $E_{\mathrm{tot}}=KIN + V$. From the mean value of the kinetic energy we obtained the parameter $\Gamma =3 N /(2\langle KIN \rangle)$.
\begin{figure}[!b]
\centering
\begin{tabular}{ c c c  }
\includegraphics[scale=0.30]{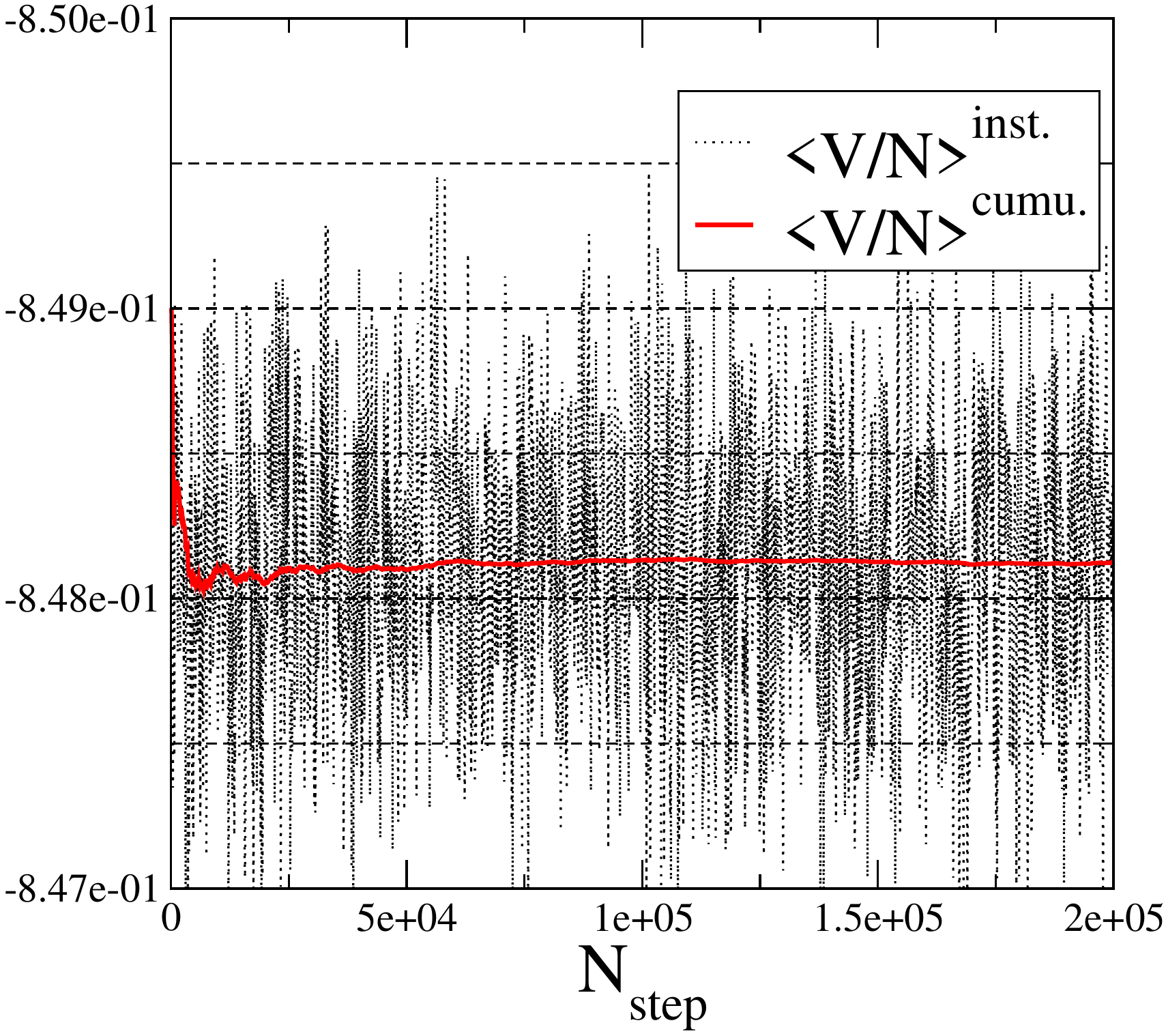} &\hspace{1.0em} & \includegraphics[scale=0.30]{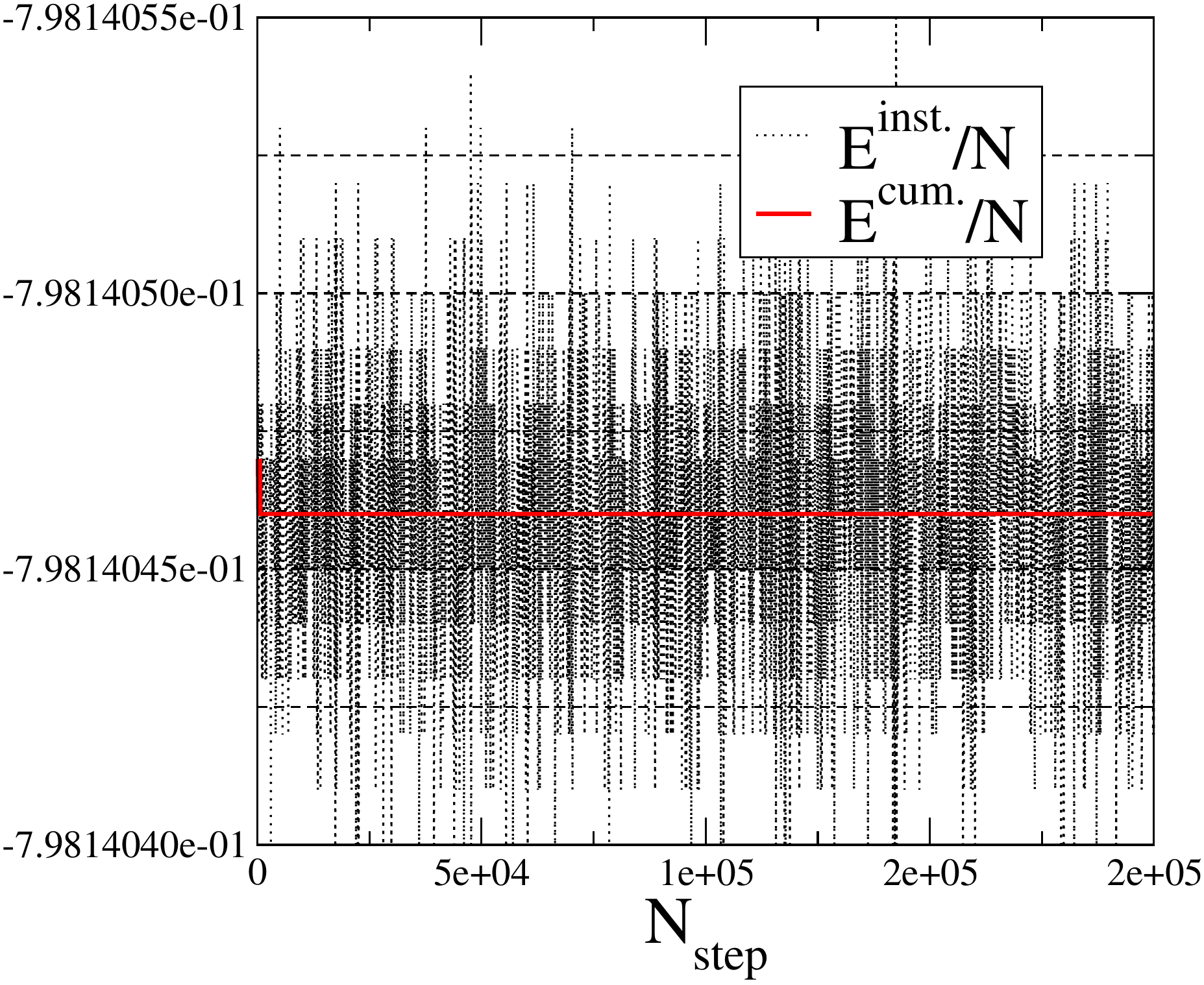} \\
\end{tabular}
\caption{(Colour online) MD simulations of the OCP with ROLL-2 algorithm. $N=2000$ ions,  time increment $h=0.01$ in reduced units. Instantaneous and cumulated potential (left-hand) and total (right-hand) energies per particle for a sub-run of $2 \times 10^5$ steps. From the mean kinetic energy one has $\Gamma \sim 30.$}
\label{FigII}
\end{figure}
 Here we present only an example to illustrate the validity of  algorithm ROLL-2. A more extensive numerical study will be published elsewhere. A sample of $N=2000$ particles was prepared in the state $\Gamma \sim 30$.  We chose for the time increment $h=0.01$ (in reduced units). The initial state was prepared so that  each of the $6$ components of  the total angular momentum bivector per particle was $\sim 10^{-16}$. ROLL-2 ensures a \emph{perfect conservation} of the angular velocity. In this respect ROLL-1 is slightly inferior because the angular velocities must be computed from the positions and the momenta yielding a slight loss in precision.
Figure~\ref{FigII} shows data for the potential and total energies for a sub-run of $2\times 10^{5}$ time steps. Since ROLL-2 is symplectic, then the conservation of the total energy is excellent.
For a complete MD simulation involving  $7\times 10^6$ time step, we obtained
\begin{itemize}
\item $\Gamma = 30.002(4)$. The reported statistical error, which corresponds to two standard deviations,   was computed by dividing the run in $n_\text{b}=1400$ blocks of $5000$ time steps and was obtained by a standard block analysis~\cite{block}.
\item $ \langle V\rangle /N= -0.848137(5)$, {i.e.}, a relative error of $6 \times 10^{-6}$. This value should be compared with the thermodynamic value $\langle V\rangle/N=-0.8481197(42)$ obtained in reference~\cite{Caillol1,Caillol2} by a finite size studies of MC simulations of the OCP at $\Gamma=30$.
\item $E/N = -0.7981404542(4)$, {i.e.}, a relative error of $5 \times 10^{-10}$. This error is surely strongly dependent on the value of the time increment $h=0.01$. Decreasing $h$ to improve on the conservation of the total energy is not a good idea since the phase space is not properly sampled. A more appealing analysis is probably to keep track of the extreme values of $E/N$ during the run. We found $-0.79814057 < E/N <  -0.79814033$ exemplifying the absence of drift and the excellent energy conservation for a quite long run of $7\times 10^6$ time steps.
\end{itemize}
\section{Conclusion}
\label{Conclusion}
In this article we have presented the symplectic integrator ROLL for MD simulations of simple fluids or plasmas on the hypersphere. It extends to hypersphere $\mathcal{S}^d$ of arbitrary dimensions, notably $d=3$, previous attempts made for 2-spheres~\cite{Vest,Lee}. The use of GA theory plays an important role in this extension because, otherwise, it would be impossible to introduce angular velocities without resorting to awkward matrix representations. In the special case $d=3$, an alternative to GA could be the use of quaternions. However,  this would be only  a peculiar case of the more general approach given here, which is also valid  for $d>3$.

Preliminary tests in the case of the OCP demonstrated the validity of ROLL for MD simulations. 
Since GA is also a very powerful approach for the classical mechanics of rigid bodies~\cite{Doran}, the extension of ROLL to molecular fluids is clearly possible. The rotations of bodies in the tangent plane $E^d_{\xi}$ could be described by the component $\Omega^{\perp}$ of the angular velocity set to zero for  point particles. Applications to polar fluids and ionic solutions is appealing.

\section*{Acknowledgements}
I am particularly grateful to  Dr. Gatien Verley for letting me discover geometric algebra and to Prof.~Alan Macdonald for useful e-mail exchanges on this subject.
I thank the editor, Prof. O. Patsahan, for her patience while waiting for the manuscript.
\appendix
\section{Geometric algebra}
\label{geometric-algebra}
We give here a brief digest on geometric algebra. It is largely inspired by the introduction to GA given by Alan Macdonald in reference~\cite{Mac-0} as well as in his textbook~\cite{Mac-1}.

The geometric algebra $\G^d$ is an extension of the inner product space $\R^d$ with more objects and operations. 
These objects are called \emph{multivectors}. They can be added or multiplied by scalars of $\R$ and these two operations confer to $\G^d$ the structure of a vector space on the field $\R$. Note that $\R$ and $\R^d$ are vector subspaces of $\G^d$.

Moreover, $\G^d$ is a graded  algebra with a product named \emph{geometric product} satisfying the properties G0-G8,  for all scalars $a$ and $A,B,C \in \G ^n$.
\begin{itemize}
\item[] {G0.} \quad There is a $1$.
\item[] {G1.} \quad $A(B+C)=AB+BC$, $(B+C)A = BA +CA$.
\item[] {G2.} \quad $(aA)B = A(aB)=a(AB)$.
\item[] {G3.} \quad $(AB)C=A(BC)$, {i.e.}, $\G^d$ is an associative but non-commutative algebra.
\item[] {G4.} \quad $1A=A1=A$. 
\item[] {G5.} \quad The geometric product of $\G^d$ is related to the scalar product of $\G^d$ by
\begin{align}
u u &= u \cdot u = \vert u \vert^2 \,
\end{align}
               \quad for all vectors $u \in \R$.
\item[] {G6.}  \quad  Every orthogonal basis $\lbrace e_i  \rbrace $, $i=1, \ldots, d$, for $\R^d$ determines a standard basis for the vector space $\G^d$. The $\binom{n}{k}$ multivectors
$e_{i_1 \ldots i_k}=e_{i_1} \ldots e_{i_k}$ with $1\leqslant i_1 < i_2  <i_k \leqslant n$ are a basis for multivectors of \emph{grade} $k$ or k-vectors. Therefore, the dimension of  $\G^d$ is $2^d$.
\item[] {G7.}  \quad Each multivector of the algebra $A$ can be uniquely expressed as a sum of k-vectors, $0 \leqslant k \leqslant n$, as  $A =\langle A \rangle_0 + \langle A \rangle_1 + \ldots \langle A \rangle_n $.
Many results of GA are an extension of the fundamental identity 
\begin{align}
\label{fonda}
u v = u \cdot v + u \wedge v  ,
\end{align}
where $u \cdot v = \langle uv \rangle_0$ is the scalar or \emph{inner} product of the two vectors
$(u,v)$. Equation~\eqref{fonda} defines their outer product as 
$u \wedge v=\langle uv \rangle_2$.
\item[] {G8.} \quad
More generally, one defines the inner and outer products of a j-vector A and a k-vector $B$ by the relations (they are non-universal and we adopt those  of Macdonald) :
\begin{subequations}
\begin{align}
A \cdot  B & = \langle  A B \rangle_{k-j}  \ , \label{inner}  \\
A \wedge B & = \langle  A B  \rangle_{k+j} \ . \label{outer}
\end{align}
\end{subequations}
\end{itemize}
We end this short and necessarily incomplete digest by listing additional definitions and useful formulae used in the main body of the paper.
\begin{itemize}
\item[] {P1.}  \quad The outer product defined at equation \eqref{outer}  is associative, which allows one to unambiguously define  the wedge product of $k$ arbitrary vectors $u_1, \ldots, u_k \in \R$. One shows that
\begin{align}
u_1 \wedge u_2 \wedge \ldots \wedge u_k & = \frac{1}{k!} \sum_{P} (-1)^{\sigma(P)} u_{P(1)} \ldots u_{P(2)}
u_{P(k)} \, , 
\end{align}
where the sum runs on all permutations $P$ of $(1, \ldots, k)$
of signature $\sigma(P)$.
 The k-vector $u_1 \wedge u_2 \wedge \ldots \wedge u_k$ is called a k-blade. For instance, $e_1 \wedge e_2 = e_1 e_2$.
To distinguish k-blades from k-vectors,  we shall denote the blades by a bold math symbol (exception 1-vectors). 
\item[] {P2.}  \quad $\mathbf{I}=e_1 e_2 \ldots e_d \equiv e_1 \wedge e_2 \wedge \ldots \wedge  e_d $ is named \emph{the} pseudo-scalar of algebra $\G^d$. Note that  $\mathbf{I}^{-1}=
e_d \wedge e_{d-1} \wedge \ldots \wedge e_1 = 
(-1)^{\frac{d(d-1)}{2}} \mathbf{I}$.
\item[] {P3.}  \quad The dual of a multivector $A$ is defined as 
\begin{align}
    A^{\star} & = A I^{-1} . \label{def-dua}
\end{align}
\item[] {P4.}  \quad If $\mathbf{A}$ is a k-blade (respect. $A$ a k-vector) of $\G^d$ then $\mathbf{A}^{\star}$ is a (n-k)blade (respect.  $A^{\star}$ a (n-k)vector) of $\G^d$. For blades, we  have a duality relation
 \begin{align}
 \mathbf{A}^{\star \star} & =\pm \mathbf{A} \label{toto}
 \end{align}
the sign depending on $d$. 
\item[] {P5.} \quad A k-blade $\mathbf{A}$ represents a vector linear subspace of $\R^d$ and its dual $\mathbf{A}^{\star}$ the orthogonal complement. An orthogonal projection of an arbitrary multivector $M$ on $\mathbf{A}$ can be expressed algebraically as 
\begin{align} \label{proj-lam}
\mathrm{P}_{\mathbf{A}}(M) &= \left(M \cdot \mathbf{A} \right) \mathbf{A}^{-1} \ .
\end{align}
\item[] {P6.}  \quad For arbitrary multivectors $A, B,C$
\begin{align}\label{util}
A \cdot ( B \cdot C) & =  (A \wedge B ) \cdot C \ .
\end{align}
\end{itemize}

\section{Rattle algorithm}
\label{b}

RATTLE numerical integration scheme  of Newton equations~\eqref{Newton_barre}  consists in the discrete mapping $(\xi_n, \dot{\xi}_n)  \mapsto (\xi_{n+1}, \dot{\xi}_{n+1})$. If applied to a particle constrained on a hypersphere, it reads: \cite{RATTLE}
\begin{subequations}\label{RAT}
\begin{eqnarray}
\xi_{n+1} &= &\xi_{n} + h \dot{\xi}_{n} + \frac{h^2}{2 m R}
           \left[  f^{\perp}(\xi_{n}) -2 \overline{\lambda}_r \xi_{n} \right]  +
           \mathcal{O}(h^3) \  \label{RAT_1},\\
 \dot{\xi}_{n+1} &=& \dot{\xi}_{n}   +\frac{h}{2 m R}  \left[ 
                    f^{\perp}(\xi_{n}) + f^{\perp}(\xi_{n+1})
                   -2 \overline{\lambda}_r\xi_{n} 
                   -2 \overline{\lambda}_s\xi_{n+1} \right]  + \mathcal{O}(h^2)\ .\label{RAT_2}
\end{eqnarray}
\end{subequations}
The two Legendre parameters $(\overline{\lambda}_r,\overline{\lambda}_s)$ are eliminated  by solving the two constraint equations
\begin{subequations}\label{RAT_C}
 \begin{align}                  
 \xi_{n+1}^2 & =  1   ,  \label{RAT_3} \\
 \xi_{n+1} \cdot \dot{\xi}_{n+1} & = 0    . 
  \label{RAT_4}               
\end{align}
\end{subequations}
In the framework of RATTLE, \emph{two} parameters $(\overline{\lambda}_r,\overline{\lambda}_s)$ are used to satisfy both position and velocity constraints~\eqref{RAT_C}; SHAKE requires only the constraint on position~\eqref{RAT_3} while the hidden constraint on the velocity~\eqref{RAT_4} is neglected. 

We first take the scalar product of both sides of equation~\eqref{RAT_1} with $\xi_n$.  Since the constraints~\eqref{RAT_C} are fulfilled at time $t_n$ it readily yields 
\begin{align}\label{lbb}
\overline{\lambda}_r &= \frac{mR}{h^2}\left(1 - \xi_{n}\cdot\xi_{n+1} \right)  .
\end{align}
Making use of expression~\eqref{lbb} in equation~\eqref{RAT_1} yields 
\begin{align}\label{eq1}
\mathrm{P}_{\xi^{\star}_n}\left( \xi_{n+1}\right) &=  h \dot{\xi}_{n} + \frac{h^2}{2 m R} f^{\perp}(\xi_{n}) + \mathcal{O}(h^3)  . 
\end{align}
Recall that \emph{the  position constraint}~\eqref{RAT_4} fully determines $\xi_{n+1}$, cf. the discussion section~\ref{ROLL-0}. For a rather small time increment $h$, one has
\begin{align}
\xi_{n+1} &= \mathrm{P}_{\xi^{\star}_n}\left( \xi_{n+1}\right) + 
\sqrt{1 -\mathrm{P}_{\xi^{\star}_n}\left( \xi_{n+1}\right)^2 } \xi_n
 .
\end{align}

As expected, $\overline{\lambda}_r$ has disappeared from equation~\eqref{eq1}
and, moreover, for a rather small $h$, $\overline{\lambda}_r$ is given by
\begin{align}\label{lambdar}
\overline{\lambda}_r &= \frac{mR}{h^2} \sqrt{1 -\mathrm{P}_{\xi^{\star}_n}\left( \xi_{n+1}\right)^2 }  .
\end{align}
With this choice for $\overline{\lambda}_r$,  the particle stays on the sphere at all times.

In a following step, making use of the expression~\eqref{lbb} of $\overline{\lambda}_r$ in equation~\eqref{RAT_2}, we obtain a simplified equation where only parameter $\overline{\lambda}_s$ survives 
\begin{align}
\label{RAT_22}
 \dot{\xi}_{n+1} &= \frac{\xi_{n+1} - \xi_n}{h} +\frac{h}{2 m R}  \left[ 
                    f^{\perp}(\xi_{n+1})
             -2 \overline{\lambda}_s\xi_{n+1}\right]  + \mathcal{O}(h^2) .
\end{align}
Taking the scalar product of both sides with vector $\xi_{n+1}$ and imposing the \emph{two constraints}~\eqref{RAT_3} and~\eqref{RAT_4}  yields
\begin{align}\label{lbbb}
\overline{\lambda}_s &= \frac{mR}{h^2}\left(1 - \xi_{n}\cdot\xi_{n+1} \right) = \overline{\lambda}_r  .
\end{align}
With this choice for $\overline{\lambda}_s$ the velocity of the particle remains tangent to the hypersphere at all times.
It is likely that the result  $\overline{\lambda}_r =\overline{\lambda}_s$ is a peculiarity of the spherical geometry. Taking  account of  expression~\eqref{lbbb} for $\overline{\lambda}_s$, we can recast equation~\eqref{RAT_22} under the form 
\begin{align}\label{e21}
\dot{\xi}_{n+1}  &=  -\frac{1}{h} \mathrm{P}_{\xi^{\star}_{n+1}}\left( \xi_{n}\right) 
 +\frac{h}{2 m R} f^{\perp}(\xi_{n+1}) + \mathcal{O}(h^2) \ , 
\end{align}
where $\mathrm{P}_{\xi^{\star}_{n+1}}\left( \xi_{n}\right)$ is the orthogonal projection
of vector $\xi_{n}$ on the hyperplane $E^{d}(\xi_{n+1})$,  perpendicular to $\xi_{n+1}$.
Finally, the RATTLE algorithm for the hypersphere $\mathcal{S}^d$ can be summarized as
\begin{subequations}\label{RATTLE-S}
\begin{align}
\mathrm{P}_{\xi^{\star}_n}\left( \xi_{n+1}\right) &=  h \dot{\xi}_{n} + \frac{h^2}{2 m R} f^{\perp}(\xi_{n}) + \mathcal{O}(h^3) \ ,  \\
\xi_{n+1} &= \mathrm{P}_{\xi^{\star}_n}\left( \xi_{n+1}\right) + \sqrt{1 -\mathrm{P}_{\xi^{\star}_n}\left( \xi_{n+1}\right)^2 } \xi_n \ , \\
\dot{\xi}_{n+1}  &=  -\frac{1}{h} \mathrm{P}_{\xi^{\star}_{n+1}}\left( \xi_{n}\right) 
 +\frac{h}{2 m R} f^{\perp}(\xi_{n+1}) + \mathcal{O}(h^2) \ . 
\end{align}
\end{subequations}
It turns out that the three equations~\eqref{RATTLE-S}
 are mathematically identical to those associated with algorithm ROLL-1, cf. equations~\eqref{ROLL-1}.
We know from the mathematical study of reference~\cite{Sympl} that RATTLE is reversible and symplectic. Therefore, the same mathematical properties hold for  ROLL.


\ukrainianpart

\title{Симплектичний інтегратор для молекулярної динаміки на гіперсфері}
\author{Ж.-М. Кайоль}
\address{
	Університет Парі-Сюд, CNRS/IN2P3,  91405 Орсе, Франція}

\makeukrtitle 

\begin{abstract}
	Ми представляємо реверсивний і симплектичний алгоритм 
	ROLL для інтегрування рівнянь руху при комп'ютерному моделюванні методом молекулярної динаміки простих плинів на гіперсфері  $\mathcal{S}^d$ довільної вимірності  $d$. Цей алгоритм виведено в рамках геометричної алгебри і показано, що він математично еквівалентний алгоритмові  RATTLE. Коротко обговорюється застосування  до моделювання методом молекулярної динаміки однокомпонентної плазми.
	
	\keywords класична статистична механіка, класичні плини, молекулярна динаміка
	
\end{abstract}
\lastpage
\end{document}